\documentclass[aps,pra,amsmath,showpacs,floatfix,superscriptaddress, twocolumn]{revtex4}
\usepackage{bm}
\usepackage{graphicx}

\begin{document}
\title{Linear optical scheme for producing polarization-entangled NOON States}

\author{Su-Yong Lee}
\affiliation{Department of Physics, Texas A\&M University at Qatar,
 Education City, POBox 23874, Doha, Qatar}

\author{Tomasz Paterek}
\affiliation{Center for Quantum Technologies, National University of Singapore, 3 Science Drive 2, 117543 Singapore, Singapore}

\author{Hee Su Park}
\affiliation{Korea Research Institute of Standards and Science (KRISS),
1 Doryong-dong, Daejeon 305-340, Republic of Korea }

\author{Hyunchul Nha}
\affiliation{Department of Physics, Texas A\&M University at Qatar,
 Education City, POBox 23874, Doha, Qatar}
\affiliation{Institute f{\"u}r Quantenphysik, Universit{\"a}t Ulm, D-89069 Ulm, Germany}

\date{\today}

\begin{abstract}
We propose a linear optical scheme that can conditionally generate high NOON states using polarization modes.
This scheme provides advantages over the previous proposals on path-entangled NOON states in view of success probability or required resources of optical elements.
We also investigate two experimental schemes feasible within existing technology that can produce the NOON-like or the NOON state for $N=4$.
\end{abstract}

\pacs{03.65.Ud, 42.50.Dv}
\maketitle

\section{Introduction}
Entanglement in multiphoton states plays a crucial role in quantum information processing \cite{HORODECKIS}.
In particular, it is known that the NOON state, i.e., the superposition of all $N$ photons in one mode $|N0\rangle$ or another $|0N\rangle$, is useful for quantum lithography and quantum metrology \cite{Dowling}.
In Mach-Zehnder-like interferometer the NOON states improve the Rayleigh diffraction limit by a factor of $N$, as the relevant signal from the NOON state spatially oscillates $N$ times as fast as the signal from a coherent state.
This effect, called super-resolution, is the basis of quantum lithography \cite{Boto}, and also leads to super-sensitivity, i.e. ability to beat the shot-noise limit in an optical phase measurement \cite{Dowling}.

Most of the NOON states proposed theoretically and produced experimentally involve superposition of different propagation directions, i.e. all $N$ photons coherently take one or another path, which is referred to as path-entangled NOON states.
A prominent example is the $2002$ state that naturally emerges at the output of the balanced beam splitter whose both inputs are supplied with single photons.
This has been known for long as the Hong-Ou-Mandel effect \cite{HOM} and the $2002$ state is as such produced deterministically.
In general, however, higher NOON states cannot be generated deterministically by simple optical tools, particularly, linear optics using phase shifters
and beam splitters \cite{Kok0,VanMeter}.
Within this paradigm, therefore, various probabilistic schemes realizing effective nonlinearity based on postselection were proposed to create $4004$ state \cite{Hwang, Zou, Fiurasek, Hofmann}, higher NOON, and NOON-like states \cite{Kok, Pryde, Sun, Kapale, Cable, McCusker, Nielsen, Hofmann1, VanMeter}.
Up-to-five-photon NOON states have been experimentally realized \cite{Mitchell, Kim, Afek}, and the super-resolution based on NOON-like states has also been demonstrated \cite{Walther, Nagata,Sun1, Resch, Liu}.

In this paper, we propose a novel scheme to generate high NOON states with two orthogonal polarization modes.
A polarization-entangled NOON state is also known to be useful for spin squeezing \cite{Shalm}. This class of NOON states has been studied in Ref. \cite{Hofmann} but the scheme there exhibits quite low success probability $P_S$, e.g., the state $4004$ is generated with $P_S=\frac{3}{256}$.
The highest known probability to generate the $4004$ state in path-entangled states is $P_S=\frac{3}{16}$ \cite{Fiurasek} that can be achieved with the resources like four single photon sources, four beam splitters, and two photon counters.
On the other hand, our scheme produces a polarization-entangled NOON state rather compactly with success probability $\frac{3}{16}$ using only one photon-counter. The reduction in the number of required photocounters is particularly desirable because the conditioning on non-detection events \cite{Fiurasek} can be significantly affected by a nonideal efficiency. We first address our proposal for the generation of $4004$ state and then its extension to $8008$ state, etc. We also investigate two practical schemes to generate the NOON-like or the NOON state for $N=4$ under realistic conditions, which seem to be very feasible within current technology.

\section{Linear optical scheme}

Consider two orthogonal polarization modes with their field operators $\hat a_H$ and $\hat a_V$ describing the horizontal and the vertical polarization, respectively.
The basic idea of our proposal is encapsulated by the following elementary identities,
\begin{eqnarray}
(\hat a_H + \hat a_V)(\hat a_H - \hat a_V) & = & \hat a_H^2 - \hat a_V^2,\nonumber\\
(\hat a_H + i \hat a_V)(\hat a_H - i \hat a_V) & = & \hat a_H^2 + \hat a_V^2,\nonumber\\
(\hat a_H^2 + \hat a_V^2)(\hat a_H^2 - \hat a_V^2) & = & \hat a_H^4 - \hat a_V^4,\nonumber\\
(\hat a_H^4 + \hat a_V^4)(\hat a_H^4 - \hat a_V^4) & = & \hat a_H^8 - \hat a_V^8.
\label{MATHS}
\end{eqnarray}
as will be addressed below.

The superpositions of field operators in the first order produce the polarization states as
\begin{eqnarray}
&& |D\rangle =\hat{a}^{\dag}_D|0\rangle=\tfrac{1}{\sqrt{2}}(\hat{a}^{\dag}_H+\hat{a}^{\dag}_V)|0\rangle=\tfrac{1}{\sqrt{2}}(|H\rangle+|V\rangle), \nonumber\\
&& |A\rangle=\hat{a}^{\dag}_A|0\rangle=\tfrac{1}{\sqrt{2}}(\hat{a}^{\dag}_H-\hat{a}^{\dag}_V)|0\rangle=\tfrac{1}{\sqrt{2}}(|H\rangle-|V\rangle), \nonumber\\
&& |L\rangle=\hat{a}^{\dag}_L|0\rangle=\tfrac{1}{\sqrt{2}}(\hat{a}^{\dag}_H+i\hat{a}^{\dag}_V)|0\rangle=\tfrac{1}{\sqrt{2}}(|H\rangle+i|V\rangle), \nonumber\\
&& |R\rangle=\hat{a}^{\dag}_R|0\rangle=\tfrac{1}{\sqrt{2}}(\hat{a}^{\dag}_H-i\hat{a}^{\dag}_V)|0\rangle=\tfrac{1}{\sqrt{2}}(|H\rangle-i|V\rangle), \nonumber
\end{eqnarray}
where $|D\rangle$ ($|A\rangle$) denotes linearly polarized state at $45$ ($-45$) degree, and $|L\rangle$ ($|R\rangle$) the left (right)-handed circularly polarized state.

\begin{figure}
\centerline{\scalebox{0.35}{\includegraphics[angle=90]{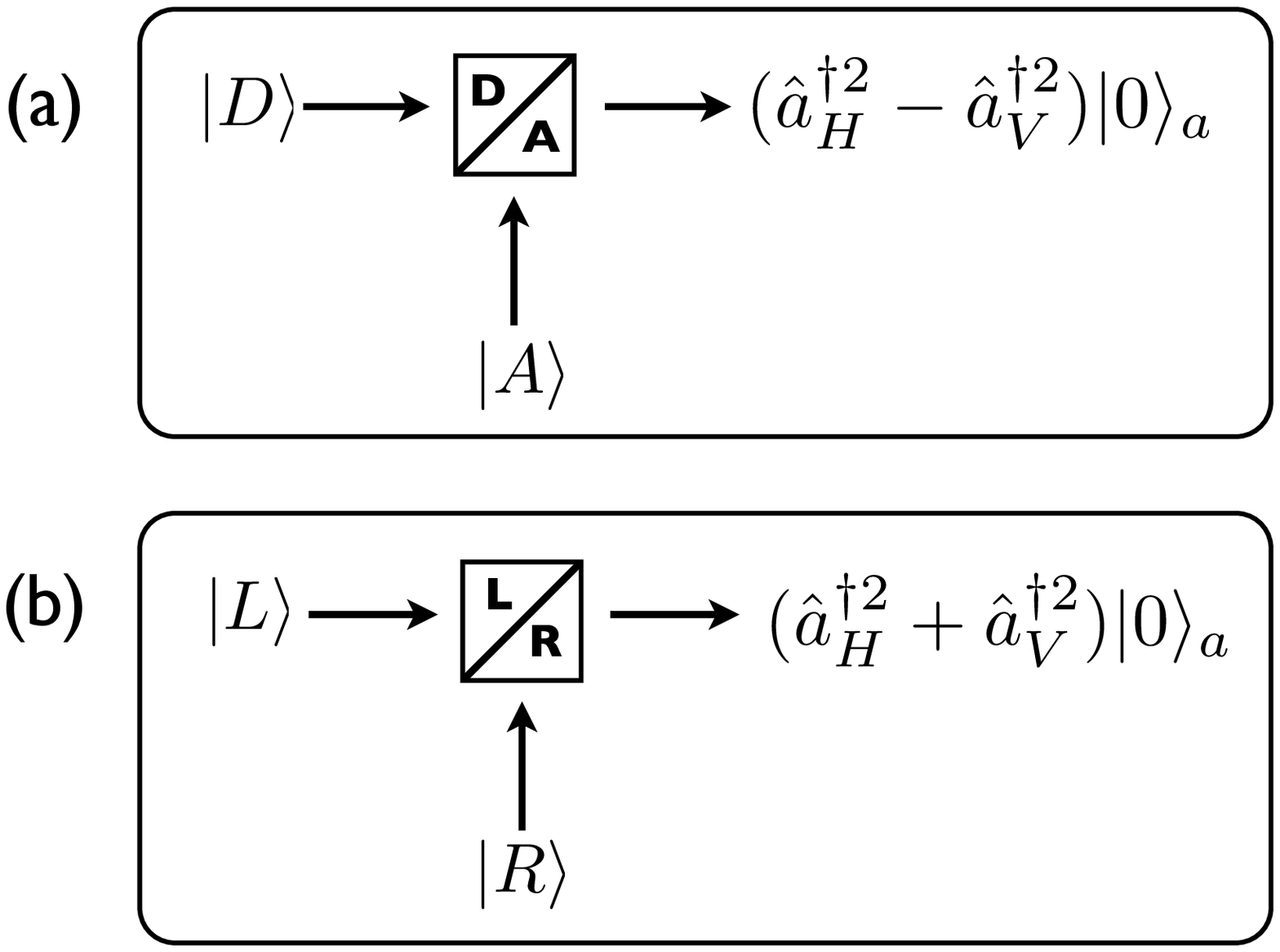}}}
\vspace{-0.8in}
\caption{Polarization analog of the Hong-Ou-Mandel (HOM) effect \cite{HOM}.
In the original HOM interferometer, the $2002$ path-entangled state is produced at the output of the balanced beam splitter whose both input ports are supplied with single photons.
Here two photons in orthogonal polarization modes are injected into two input ports of a polarizing beam splitter that transmits one polarization and reflects the other.
If the input photons are (a) linearly polarized along $+45^{\circ}$ ($|D\rangle$) and $-45^{\circ}$ ($|A\rangle$), or (b) left-circularly ($|L\rangle$) and right-circularly ($|R\rangle$) polarized, the resulting state is a $2002$ state in the horizontal/vertical basis.}
\label{FIG_2002}
\end{figure}

On the other hand, the superpositions of the field operators in the second order like the ones in Eq. (\ref{MATHS}) can be implemented using polarizing beam splitters (PBSs).
PBS is a device that transmits one polarization light and reflects the other polarization orthogonal to the transmitted one.
For example, the state $\hat{a}^{\dag}_D\hat{a}^{\dag}_A|0\rangle=\frac{1}{2}(\hat{a}^{\dag 2}_H - \hat{a}^{\dag2}_V)|0\rangle$ emerges at the output of the PBS that transmits $|D\rangle$ and reflects $|A\rangle$ as shown in Fig. \ref{FIG_2002}.
Similarly, the state $\hat{a}^{\dag}_L\hat{a}^{\dag}_R|0\rangle=\frac{1}{2}(\hat{a}^{\dag 2}_H + \hat{a}^{\dag2}_V)|0\rangle$ is obtained via the PBS that transmits $|L\rangle$ and reflects $|R\rangle$.
Note that both of these polarization $2002$ states are produced deterministically in analogy to the Hong-Ou-Mandel effect.

One may consider an extension of the device described above that could produce a $4004$ state deterministically similar to the case of $2002$ state.
Based on the above-mentioned method, one can separately prepare two $2002$ states $|\Psi\rangle_a=\hat a_D^\dag \hat a_A^\dag |0\rangle=\frac{1}{2}\left(\hat a_H^{\dag2}-\hat a_V^{\dag2}\right)|0\rangle$ and
$|\Psi\rangle_b=\hat b_L^\dag \hat b_R^\dag |0\rangle=\tfrac{1}{2}(\hat b_D^{\dag2}+\hat b_A^{\dag2})|0\rangle=\tfrac{1}{2}(\hat b_H^{\dag2}+\hat b_V^{\dag2})|0\rangle$. Now one may inject these two states $|\Psi\rangle_a$ and $|\Psi\rangle_b$ as inputs to a PBS in order to deterministically produce the $4004$ state at the output, which can be proven impossible within the regime of linear passive transformations from input to output modes \cite{VanMeter}.
Thus, let us consider a scheme where a $4004$ state can be conditionally generated using linear optics.
We describe the input modes of the beam splitter by operators $\hat a$ and $\hat b$.
These input ports are supplied with two pairs of $2002$ states obtained as before, namely the input state is
$\hat{a}^{\dag}_D\hat{a}^{\dag}_A\hat{b}^{\dag}_L\hat{b}^{\dag}_R|0\rangle=\frac{1}{4}(\hat{a}^{\dag 2}_H-\hat{a}^{\dag 2}_V)
(\hat{b}^{\dag 2}_H+\hat{b}^{\dag 2}_V)|0\rangle$.
The beam splitter transforms the initial modes as $\hat a \to t \hat a + r \hat b$ and $\hat b \to t \hat b - r \hat a$,
where $t$ and $r$ denote transmissivity and reflectivity, respectively, of the beam splitter.
This transformation yields the state $(t\hat{a}^{\dag}_D+r\hat{b}^{\dag}_D)(t\hat{a}^{\dag}_A+r\hat{b}^{\dag}_A)(t\hat{b}^{\dag}_L-r\hat{a}^{\dag}_L)(t\hat{b}^{\dag}_R-r\hat{a}^{\dag}_R)|0\rangle$.
Under the non-detection event in the output mode $\hat b$, the state of the other output mode becomes
\begin{eqnarray}
t^2r^2\hat{a}^{\dag}_D\hat{a}^{\dag}_A\hat{a}^{\dag}_L\hat{a}^{\dag}_R|0\rangle
&=&\frac{t^2r^2}{4}(\hat{a}^{\dag 4}_H-\hat{a}^{\dag 4}_V)|0\rangle, \\
&=&\sqrt{3}t^2r^2\frac{(|4\rangle_H|0\rangle_V-|0\rangle_H|4\rangle_V)}{\sqrt{2}}. \nonumber
\end{eqnarray}
This is the very $4004$ state with the unnormalized weight representing the success probability for its generation, $P_{4004} = 3 t^4 r^4$.
This probability is optimized by a balanced beam splitter $t^2 = r^2 = \tfrac{1}{2}$, for which we obtain $P_{4004} = \tfrac{3}{16}$.
Using single photon sources, linear optics, and photon counters, previous schemes have shown the success probability of $\tfrac{3}{64}$ \cite{Hwang}, $\tfrac{1}{6}$ \cite{Zou}, and $\frac{3}{16}$ \cite{Fiurasek}. In particular, compared with the Fiur\'a\u sek's protocol \cite{Fiurasek}, where four beam splitters and two photon counters are required,
our scheme rather compactly employs two PBS, one unpolarized BS, and only one photon counter.

As indicated in Eq. (\ref{MATHS}), a similar method can be used to generate higher NOON states.
For example, consider two identical setups each of which generates the $4004$ state in one of its output modes.
Each setup now succeeds with probability $P_{4004} = \tfrac{3}{16}$, and one of these two identical states $\tfrac{1}{\sqrt{2}}(|40\rangle - |04\rangle)$ can be transformed into another state $\tfrac{1}{\sqrt{2}}(|40\rangle + |04\rangle)$ by putting a $\tfrac{\lambda}{8}$-wave plate in the optical path.
This wave-plate can be achieved by a combination of standard quarter-wave and half-wave plates or with the help of birefringent photonic crystal.
Injecting states $\tfrac{1}{\sqrt{2}}(|40\rangle \pm |04\rangle)$ into different input ports of the balanced beam splitter generates the $8008$ state with success probability $P_{8008} \approx 0.005$.
Finally, note that the NOON states with $N \le 7$ photons can be obtained by photon subtraction operations performed on the $8008$ state.
Specifically, when a coherent operation of photon subtraction $\hat{a}_H+\hat{a}_V$ \cite{Grangier} is applied to a $N00N$ state $|N,0\rangle + |0,N\rangle$, the output state becomes $|N-1,0\rangle + |0,N-1\rangle.$ In the case that two polarization modes $H$ and $V$ take the same path in the ouput, the operation $\hat{a}_H+\hat{a}_V$ is simply the photon subtraction for a single mode linearly polarized along $+45^{\circ}$. As an illustration, we plot in Fig.2 the fidelity of output state by applying the photon-subtraction operation $\hat{a}_1+\hat{a}_2$ to an ideal $|N,0\rangle + |0,N\rangle$ using an on-off photodetector with efficiency $\eta$. We see that the high fidelity is maintained even with a very low efficiency $\eta$, which implies that the only practical key issue is the generation of the initial NOON state with high fidelity.

\begin{figure}
\centerline{\scalebox{0.35}{\includegraphics[angle=270]{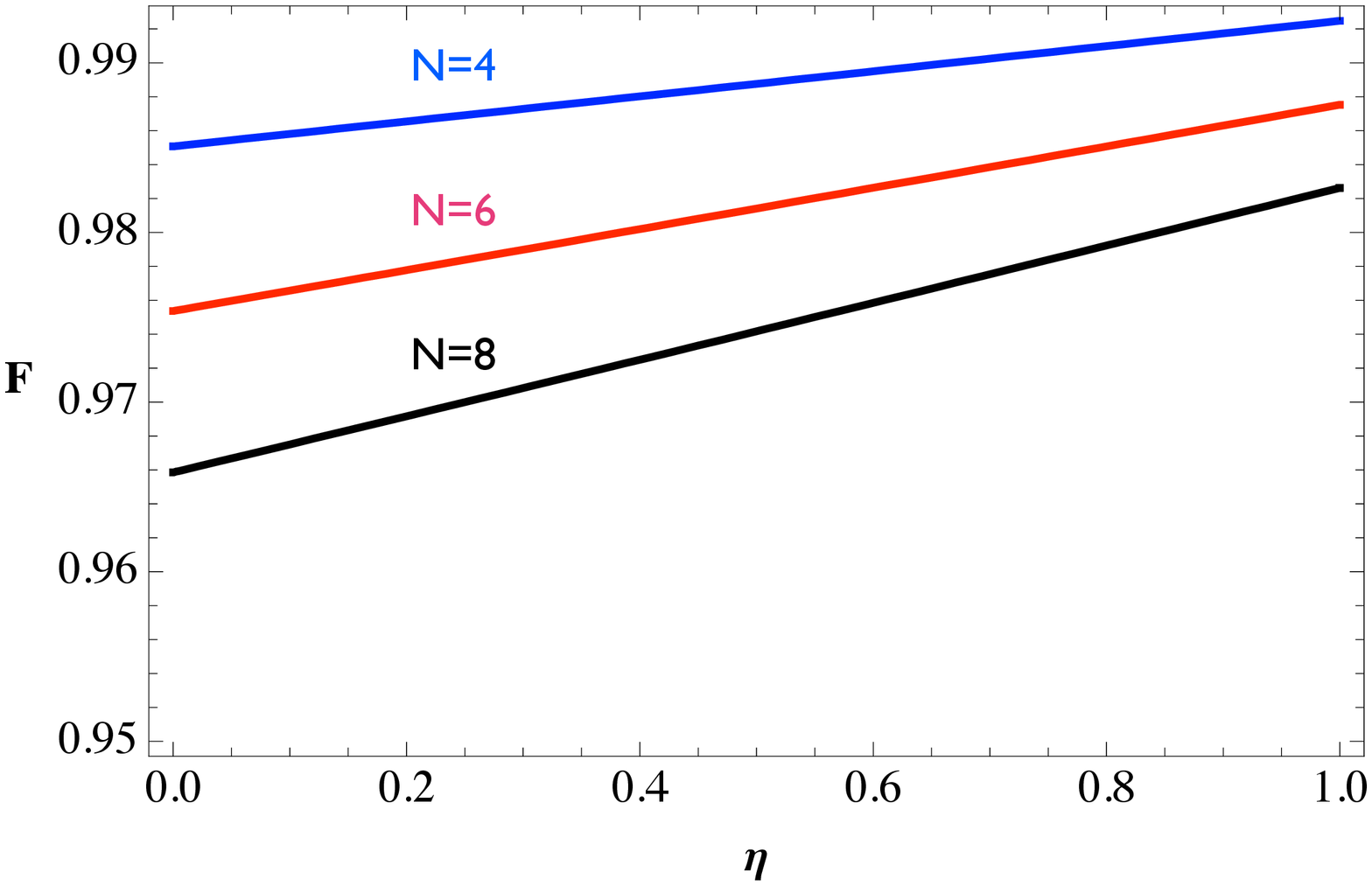}}}
\vspace{-0.3in}
\caption{The fidelity of the output state by applying the photon-subtraction operation $\hat{a}_1+\hat{a}_2$ to an ideal $|N,0\rangle + |0,N\rangle$ using an on-off photodetector with efficiency $\eta$. We used the value of 0.99 as the beam-splitter transmittance to implement the coherent photon subtraction $\hat{a}_1+\hat{a}_2$.}
\label{subtraction}
\end{figure}

\section{Realistic schemes}

\begin{figure}
\centerline{\scalebox{0.4}{\includegraphics[angle=90]{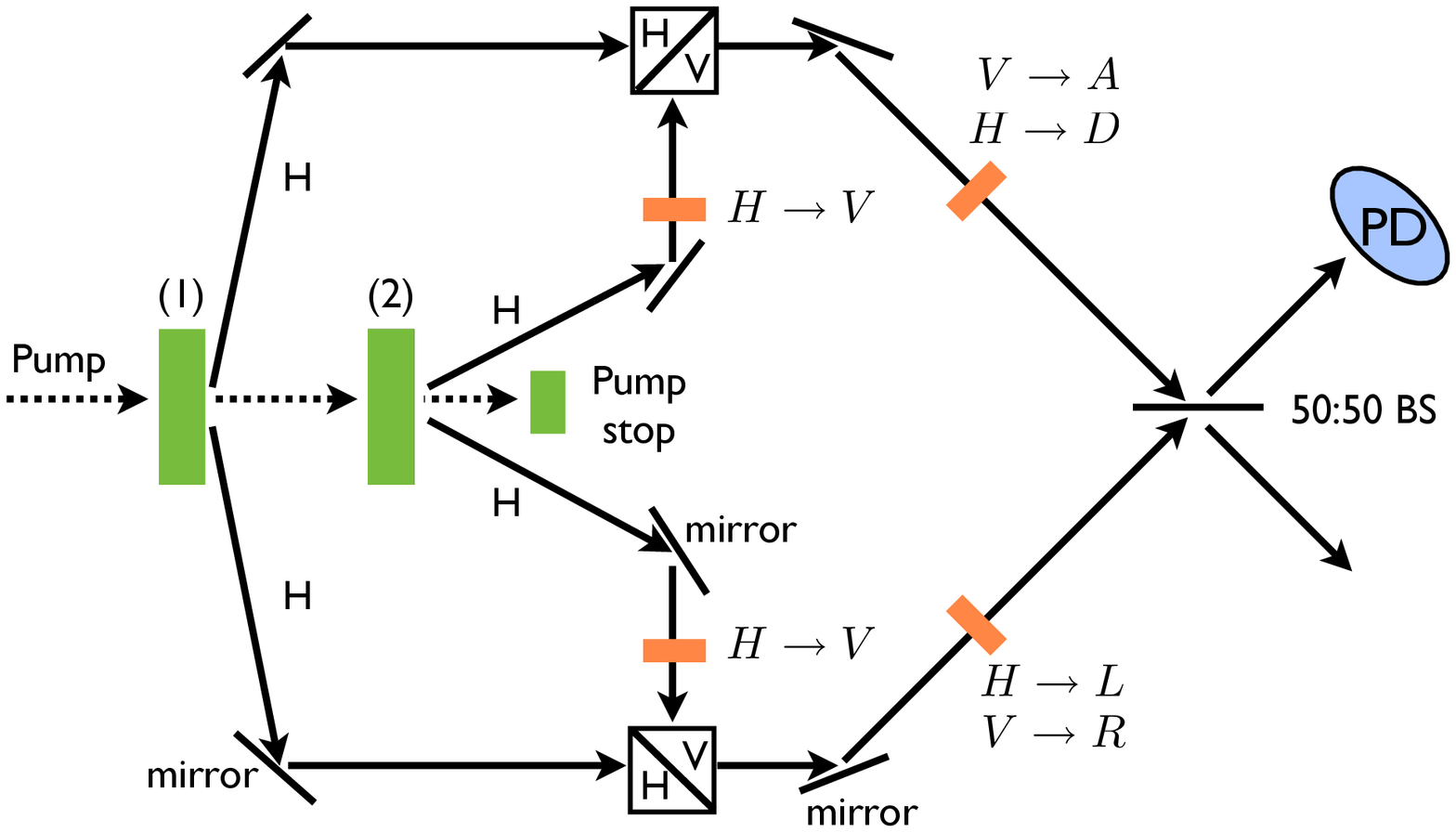}}}
\vspace{-1.2in}
\caption{A practical scheme to produce a NOON-like state for $N=4$.}
\label{FIG_SETUP}
\end{figure}

In this section we consider two practical schemes to produce the NOON or the NOON-like states under realistic experimental conditions.
The first setup is depicted in Fig. \ref{FIG_SETUP}, where the four photons are generated via spontaneous parametric down-conversions using two non-linear crystals.
Let us consider the process in which the pump field produces a pair of photons with the same polarization, say $|H\rangle$.
Four photons will be produced via two pair emissions, which can occur in three different ways:
(i) Both pairs are generated in the first crystal,
(ii) both pairs are generated in the second crystal, and
(iii) one pair generated in each crystal.
Since the production rate of two pairs in a single crystal is the same as that of two pairs in different crystals \cite{Nagata},
we consider an equal superposition of these three cases as an initial state.
This lack of which-path information leads to a $4004$-like state instead of the $4004$ state, as will be shown below.
We also note that the role of the two crystals can be replaced by a single crystal in a configuration of Walther \emph{et al.} \cite{Walther}

In Fig. 3, two pairs of photons are emitted via spontaneous parametric down-conversion in crystals (1) and (2).
Taking the initial polarization of the down-converted photons to be $|H\rangle$, we rotate it to $|V\rangle$ using polarization rotators denoted by $H \to V$ for all photons produced in the second crystal.
Next, we superpose photons from both crystals on two PBSs that transmit polarization $|H\rangle$ and reflect $|V\rangle$.
Photons generated this way travel towards a balanced beam splitter via the upper path $a$ and the lower path $b$.
The polarization of the photons in mode $a$ is additionally rotated as $H \to D$ and $V \to A$,
whereas those in mode $b$ as $H \to L$ and $V \to R$.
After the final beam splitter, the output state is given by
\begin{eqnarray}
&&\frac{1}{16\sqrt{3}}[2(\hat{a}^{\dag 4}_H-\hat{a}^{\dag 4}_V)+2(\hat{b}^{\dag 4}_H-\hat{b}^{\dag 4}_V)
-4(\hat{a}^{\dag 2}_H\hat{b}^{\dag 2}_H-\hat{a}^{\dag 2}_V\hat{b}^{\dag 2}_V)\nonumber\\
&&+4i(\hat{b}^{\dag 2}_H-\hat{a}^{\dag 2}_H)(\hat{b}^{\dag 2}_V-\hat{a}^{\dag 2}_V)]|00\rangle_{ab},
\end{eqnarray}
which, after post-selection of no photons in photo-detector PD \cite{note}, becomes
\begin{eqnarray}
\frac{1}{16\sqrt{3}}[2(\hat{a}^{\dag 4}_H-\hat{a}^{\dag 4}_V)+4i\hat{a}^{\dag 2}_H\hat{a}^{\dag 2}_V]|0\rangle_a.
\label{400422}
\end{eqnarray}
This state is a superposition of the $4004$ state and another state in which two photons are $|H\rangle$ polarized and two photons $|V\rangle$ polarized.
The NOON-like state has been utilized to demonstrate super-sense interferometry beating the shot-noise limit \cite{Nagata}, and
for metrology purpose, further polarization rotator acting as $H \to D$ is required together with photon number resolving detectors.

A perfect $4004$ state can be obtained if in place of down-conversion crystals one uses heralded source of single-photons \cite{PAN_HERALDED,WALTHER_HERALDED}.
We now consider another scheme to employ the heralded single photons and investigate the effect of nonideal photodetector on the quality of produced state.
Specifically, we consider a realistic scheme with the use of on-off detector for the conditioning on non-detection events.
The on-off detector can be described by a two component positive-operator valued measure (POVM) \cite{Mog, Rossi},
$\hat{\Pi}_0=\sum_n (1-\eta)^n|n\rangle\langle n|$ (no click), and $\hat{\Pi}_1=\hat{I}-\hat{\Pi}_0$ (click), where $\eta$ is the detector efficiency.
After the non-detection event at PD, we obtain the fidelity of the output state
, $F=\langle 4004|\rho_{out}|4004\rangle=\frac{3}{(2-\eta)^2(4-4\eta+3\eta^2)}$, compared with the ideal $4004$ state. Using this practical state, we also consider the phase sensitivity $\Delta\varphi$  in a Mach-Zehnder interferometer by parity measurement \cite{Bollinger,Gerry,Aravind, Plick1, Plick2}, which is given by
\begin{eqnarray}
\Delta \varphi  &=& \frac{1}{4}\frac{\sqrt{1/F^2-[(1-\eta)^4-\cos(4\varphi)]^2}}{\sin(4\varphi)} \nonumber\\
&\geq& \frac{1}{4}\sqrt{1/F^2-(1-\eta)^8}.
\end{eqnarray}
To overcome the shot-noise limit $1/2$ for the case of average photon number $4$, the detection efficiency should be larger than $0.519$, as shown in Fig. 4.
It can be achieved with the current technology, as the on-off detection efficiency has attained up to $66\%$ \cite{Brida}.
In this regime, we also obtain the fidelity over $0.5$. In other words, the fidelity above $50\%$ is sufficient for the usefulness in the phase sensitivity.

\begin{figure}
\centerline{\scalebox{0.7}{\includegraphics{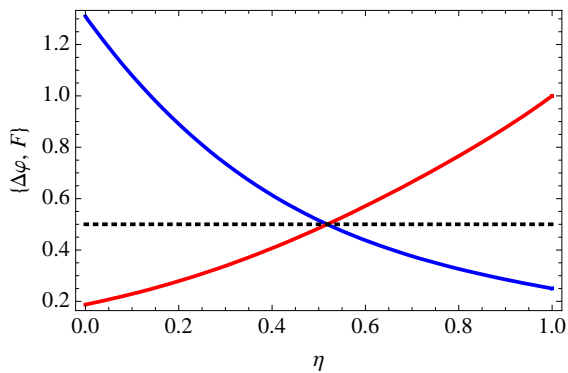}}}
\caption{Phase sensitivity (blue) in the Mach-Zehnder interferometer and fidelity (red) for the practically produced $4004$ state as a function of detection efficiency $\eta$.
Black-dotted line is the shot-noise limit $1/2$ for the case of $N=4$.}
\label{FIG_PHASE}
\end{figure}

\section{conclusion}
In this paper, we have proposed a linear optical scheme that can conditionally generate a polarization-entangled NOON state with the highest known success probability. This scheme is rather compact compared with the previous proposals, particularly with the use of only one photo-detector for a conditional implementation.
The scheme can be extended to higher NOON states and we have also investigated two practical schemes that can generate the NOON-like or the NOON-state for $N=4$ under realistic conditions.

\begin{acknowledgments}
This work is supported by the NPRP 08-043-1-011 from Qatar National Research Fund.
TP acknowledges financial support by the National Research Foundation and the Ministry of Education in Singapore, and
HN the support by the Alexander von Humboldt Foundation.
\end{acknowledgments}


\begin{thebibliography}{99}
\bibitem{note} We envision here the implementation of our proposed scheme in a pulsed regime. Namely, a strong pulse to pump the nonlinear crystals is generated at specified times in repetition, then the experimenter can know in advance when the output signal would arrive at the photodetector, together with the information on the optical path of the whole experimental setting.
One may use the output state for a further processing like phase-measurement only if the detector does not click at the times of expected signal.
\end{thebibliography}
\end{document}